# Does normal pupil diameter differences in population underlie the color selection of the #dress?

*Kavita Vemuri\*,[1] Kulvinder Bisla,[1] SaiKrishna Mulpuru,[1] Srinivasa Varadarajan[2]*
1. International Institute of Information Technology, Hyderabad, India
2. L V Prasad Eye Institute, Hyderabad
([*kvemuri@iiit.ac.in](*kvemuri@iiit.ac.in))

The fundamental question that arises from the color composition of the #dress is: 'What are the phenomena that underlie the individual differences in colors reported given all other conditions like light and device for display being identical?'. The main color camps are blue/black (b/b) and white/gold (w/g) and a survey of 384 participants showed near equal distribution. We looked at pupil size differences in the sample population of 53 from the two groups plus a group who switched (w/g to b/b). Our results show that w/g and switch population had significantly ( w/g <b/b, p-value = 0.0086) lower pupil size than b/b camp. A standard infinity focus experiment was then conducted on 18 participants from each group to check if there is bimodality in the population and we again found statistically significant difference (w/g < b/b , p-value = 0.0132). Six participants, half from the w/g camp, were administered dilation drops that increased the pupil size by 3-4mm to check if increase in retinal illuminance will trigger a change in color in the w/g group, but the participants did not report a switch. The results suggest a population difference in normal pupil-size in the three groups.

## 1.INTRODUCTION

The debate and initial experiments to understand the reasons for individual variation in the colors people see in an over-exposed blurry photograph of a #dress has bought into spotlight the deliberation on color as perception or physiological. The #dress is stated to be originally blue/black by the designer but when viewed under identical conditions one set of the viewers reported it was white/gold (w/g), while others saw it as blue/black (b/b) and some have switched mostly from w/g to b/b. A task which required participants to color match patches from the dress showed differences between the two color camps due to lightness and not in chromaticity [1]. An online survey with a sample size of 1401 [2] indicated gender differences, with more females and older population choosing w/g; they also found that people who spent longer periods in artificial light reported seeing the color as b/b. A bias towards bluish illumination in interpretation of illuminance and color constancy was reported in another study [3]. Review [4] of the initial findings from [1,2,3] suggests that results need to be analyzed with a volume of literature on color perception from sensory inputs to the underlying neural mechanisms. Testing for variation due to perceptions of light, hue and chromaticity a large hue-angle range with points being in both gold and black was reported [5]. The studies to date confirm that there are three main population groups – the *w/g'ers'*, *b/b'ers* and the *switchers*. The question that still remains is: what could be the reason for individuals to see 2 different colors given all other viewing conditions the same?. To explore the intrigue, we investigated the pupil size variations different colors including the colors of the dress – white, gold, blue (light to dark shades), black, brown and grey. Our initial premise was based on spectral sensitivity of the eye and the role of the pupil in retinal illuminance, and hence an experiment was design to measure pupil response to color from a display to check if the RGB and the colors from the dress would show difference in either the size or response latency as could be seen when shifting from mesopic to photopic level.

The human eye's pupil diameter falls between 2 and 8 mm and determines the optical transfer function, and the amount of light that falls on the retina. The fovea in the retina is approximately 1.4 mm in diameter with rod-free areas of 0.5 mm with 0.3 mm being populated by purely by cones (L and M cones). Various factors are known to affect pupil size and pupillary light reflex, like retinal illuminance [6, 7,8], accommodative state [9], and iris color and age [10]. The pupillary light reflex is in turn controlled by photosensitive retinal ganglion cells which contain the photopigment melanopsin [11,12] and also by external inputs channeled by the rods and cones [13]. The cones are responsible for our color sense. The retinal illuminance mediating the pupil light reflex has been studied as a function of luminance, stimulus size and the wavelength (color) and estimated as a product of the corneal field density and the stimulus illuminance [14,15]. The wavelength dependency of corneal field density [8,16] studied by varying the retinal illuminnace with flashing colors indicated that pupil light reflex was stimulus illuminance and size dependent under certain incidence conditions

Extending the understanding of light, retinal illuminnace and pupil light reflex, the present study was designed to examine the color (wavelength) modulated pupil diameter changes by presented different color slides in natural viewing conditions – like on a laptop LCD screen. The pupil size variation was studied using an infrared camera based eye tracker at 30 Hz for 23 colors including RGB, black/white. Using a range of colors, we could get the variation within each color stimuli and the change over the 130 second duration of the experiment. The large data set gives a time-averaged stable pupil size for each participant.

## 2. METHODS
### A. Participants:
A survey was completed by 384 individuals (male: 290 and female:94, mean age: 22) all of Asian-Indian ethnicity with iris color in shades of brown. Fifty three participants (b/b: 21, w/g: 16 and switch: 16, mean age: 27 years) took part in the pupil size measurement experiment using a Tobii X30 eye tracker. Twenty from this subset were naïve, that is, they had not seen the image before. The artificial dilation using prescription drops was administered by an ophthalmologist on 6 participants from the above set (b/b: 3, w/g: 3, though one 'switched' just prior to the experiment). Upon comprehensive eye examination by the ophthalmologist, all subjects were found to have best corrected visual (Snellen) acuity of 20/20 or better and no visual abnormality. Informed consent was obtained from all participants and the eye tracking studies are approved by the human ethics committee of the International Institute of Information Technology, Hyderabad.

### B. Apparatus
The Tobii X30 and X120 eye trackers were used to collect pupil size data for the color slides and the infinity focus experiments respectively. The light flux emitted by the LCD screen of the HP laptop and the background light in the room were measured by a digital light meter (Lutron LX-102).

### C. Stimuli and experimental procedure
**Color Slide experiment:** 22 different color shades between blue and red were displayed in no particular color order for 3 seconds on a HP Pavilion laptop screen (15inch). Included were the hues of the dress colors (shades of white, blue, black, brown and gold) obtained by reading the pixel values from a digitized photograph of the #dress. The pixel value at different points on the dress digital photograph was acquired and these values were input into the paintbrush tool to make the color slides. To reduce color/luminance adaptation effects, a filler slide in a shade of gray adjusted using the RGB values was introduced between successive color slides for 3 seconds. The participants were seated at 60 cms from the screen and were instructed to look at the screen as normally as possible. The Tobii X30 eye tracker continually recorded the pupil size at the rate of 30 Hz while the participants looked at the color slides. A total of 43 slides were shown in a lighted room (background illumination: 168 lux) and with the lights in the room turned off (room illumination: 7 lux) and the only light being form the laptop screen. At the beginning and the end of the color slide show the #dress image was shown and the participants were asked to indicate the color composition they saw as b/b or w/g in 2 choice forced alternative manner. In addition, participants were asked if they had at any time 'switched'; those who answered 'Yes' formed the 3rd group. The subset of participants who switched, mentioned seeing a w/g the first time and then continued to see b/b, except for one who could switch at 'will'; pupil data was collected from this participant, but not presented in this report. All the participants were given 5 minutes of adaptation time before the experiment.

**Infinity focus:** A black-colored 'X' sign was projected on a grayish wall 4 meters away from a participant for 3 seconds in a well-lit room with both artificial and daylight mixed (230lux). An instruction slide at the beginning and the #dress were again shown and the choice noted down at the beginning and end while pupil size was collected using a Tobii X120. Out of the 36 participants, 24 were new and 6 took part in the previous experiment. All other procedures were as before.

**Artificial Pupil dilation:** Six participants from among the 53 went through a comprehensive eye test in an eye clinic and were also tested for color vision. A day after the eye test, the participants took part in the pupil size measurement where slides with the red, green, blue colors as shown in the first experiment plus black was preceded and followed by the dress photograph with the choice question. An ophthalmologist reviewed the reports and administered dilation drop (1 drop) in a light controlled room (background luminance: 168 lux) and the experimental color slides was repeated twice with a gap of 30 minutes between the two presentations.

## 3. RESULTS AND DISCUSSION
**Survey :** To check for age and gender dependency, a survey with 384 participants (male: 290 and female: 94) all of Asian-Indian origin with iris color in shades of brown and across age groups (10 to 60 years) was conducted. The survey participants viewed the image on their personal electronic devices like laptops, smart phones, desktop monitors and filled an online question with choices: b/b, white/gold or seen both and we found no dependency on age or gender with 48.5% reported seeing the dress in blue/black and 51.5% selecting the white/gold-brown option. With regards to gender the break up was 50% of males and 51% of female participants seeing the dress as blue and black while 49% male and 48% female report the color as white and gold/brown. Around 10% of the survey participants mentioned the ability to switch from white/gold-brown initially to blue/black but not being able to switch back

**Color slide experiment:**

Pupil size measurements while viewing colors displayed on a light-emitting LCD screen were collected from 53 respondents with the following break-up: blue/black (21), white/gold (16) and switch (16). For the analysis, 2 participants' data were discarded due to insufficient eye tracking data. To look at group wise pupil size response differences for each color the average pupil size across all subjects in each group for individual colors was plotted for both light conditions (Fig 1a). As average values across the 3 second color stimulus period was considered, pupillary unrest or *hippus* which is said to induce a low frequency random fluctuation (0.02 to 2Hz) with amplitude of approximately 0.25 [18] effects were not regarded. The average pupil diameter for the b/b group is higher across all the colors compared to the w/g or switch groups. Interestingly, the w/g and switch group had nearly equal pupil size across most of the colors. The adjacent color variation in pupil size is similar for all three groups, with small differences (0.1 to 0.2mm) between the b/b and w/g/switch for grey-red or red-grey and grey-pink-grey-orange sequences. The data collected when the room lights were turned off and the only light emitting device in the room was the laptop screen, the b/b group's pupil size shows a dc shift downwards with a maximum of 0.49 mm for 'dark_blue' and a minimum for the 'purple'; also the inter-color pupil size change for green-grey-red-grey-orange sequence was 0.2mm on an average in the dark room. Similar and significant changes were not observed for the w/g or switch groups except for the 'grey_blue' (0.2mm). The pupil size changes for the RGB and white/black colors adhere to expected gradients with 'black' screen inducing a pupil size increase and bright 'white' forced a reflexive size decrease. The change noticed in the pupil size to color is in accordance to the way the brain compares the responses from cones types. For example, in the log luminance versus pupil diameter plot (Figure 1b,c) we notice constriction of the pupil for 'green' and 'yellow' and dilation for 'black'. The theoretical pupil diameter calculated using the equation: pupil size = $4.9 - 3\tanh(0.4 \log L)$, where L is luminance measured by the light meter (Lx$10^2$) at 2inch from the screen for each color [19] is also included in the plots for lighted (Figure 1b) and dark room (Figure 1c) conditions. The statistical significance was estimated by a Wilcoxon-Mann-Whitney Test, with the condition: w/g <b/b , with inputs as the average pupil size across all colors for each member of the group and the p-value was 0.0086 (for lighted room) and 0.0391 (for dark room) indicative of significant difference especially for the lighted room condition between the two groups in pupil size.

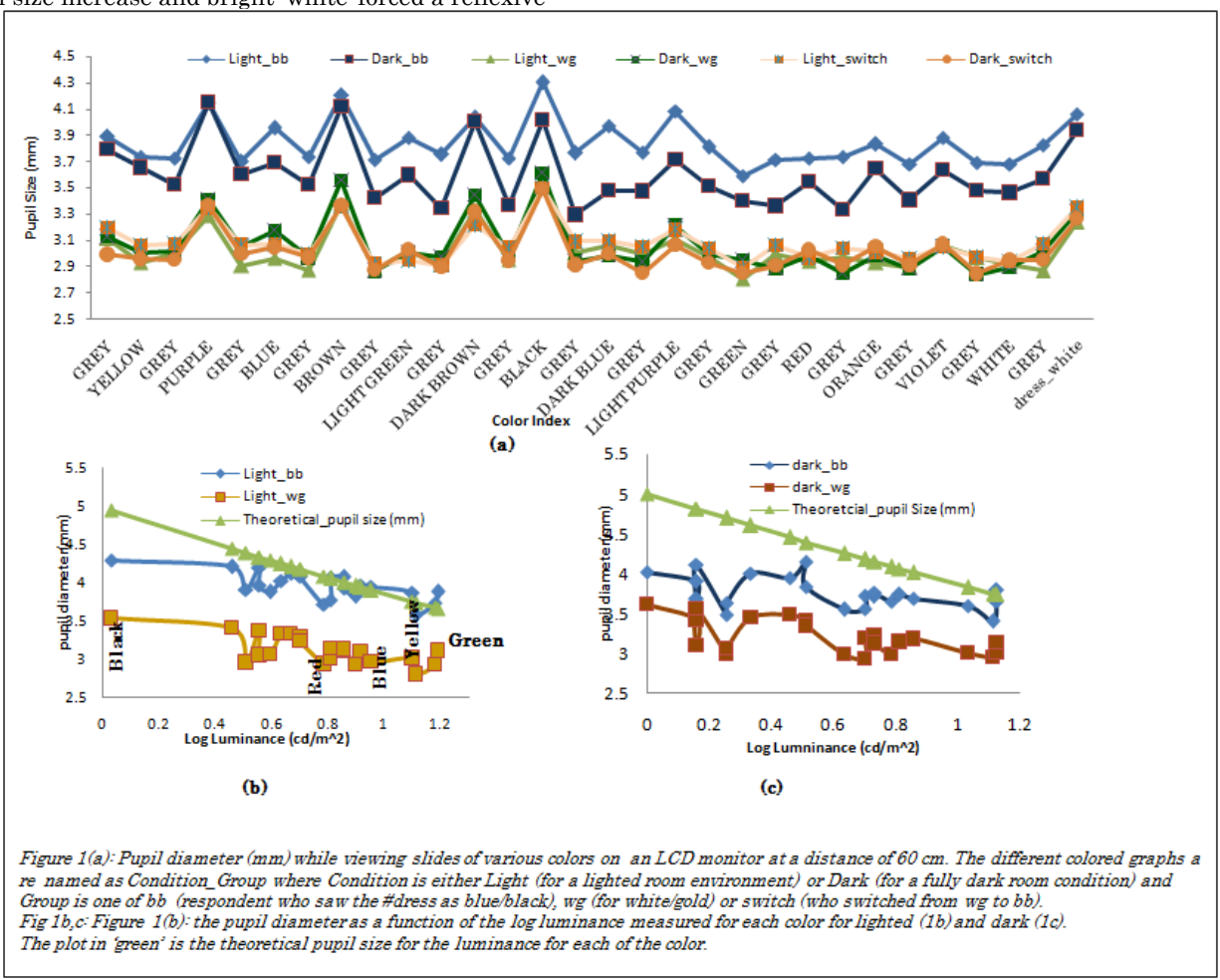

Figure 1(a): Pupil diameter (mm) while viewing slides of various colors on an LCD monitor at a distance of 60 cm. The different colored graphs are named as Condition_Group where Condition is either Light (for a lighted room environment) or Dark (for a fully dark room condition) and Group is one of bb (respondent who saw the #dress as blue/black), wg (for white/gold) or switch (who switched from wg to bb).
Fig 1b,c: Figure 1(b): the pupil diameter as a function of the log luminance measured for each color for lighted (1b) and dark (1c). The plot in 'green' is the theoretical pupil size for the luminance for each of the color.

Considering that the switch and w/g group show similar values we plot the average pupil size for left/right eye estimated from 90 data points from seven colors: black, blue, green, red, grey, yellow, white for individual participant for the b/b and w/g groups in the lighted room condition (the trend is similar for dark room data too but shifted down by 2-3mm for the b/b group). A 2-tailed unequal variance T-Test was also committed for each color across the 2 groups, and the p-values for each of the 7 main colors are listed in the table 1. The within-subject size variation for the colors is substantial as expected from color perception even at constant light intensity of the emitting surface (laptop).

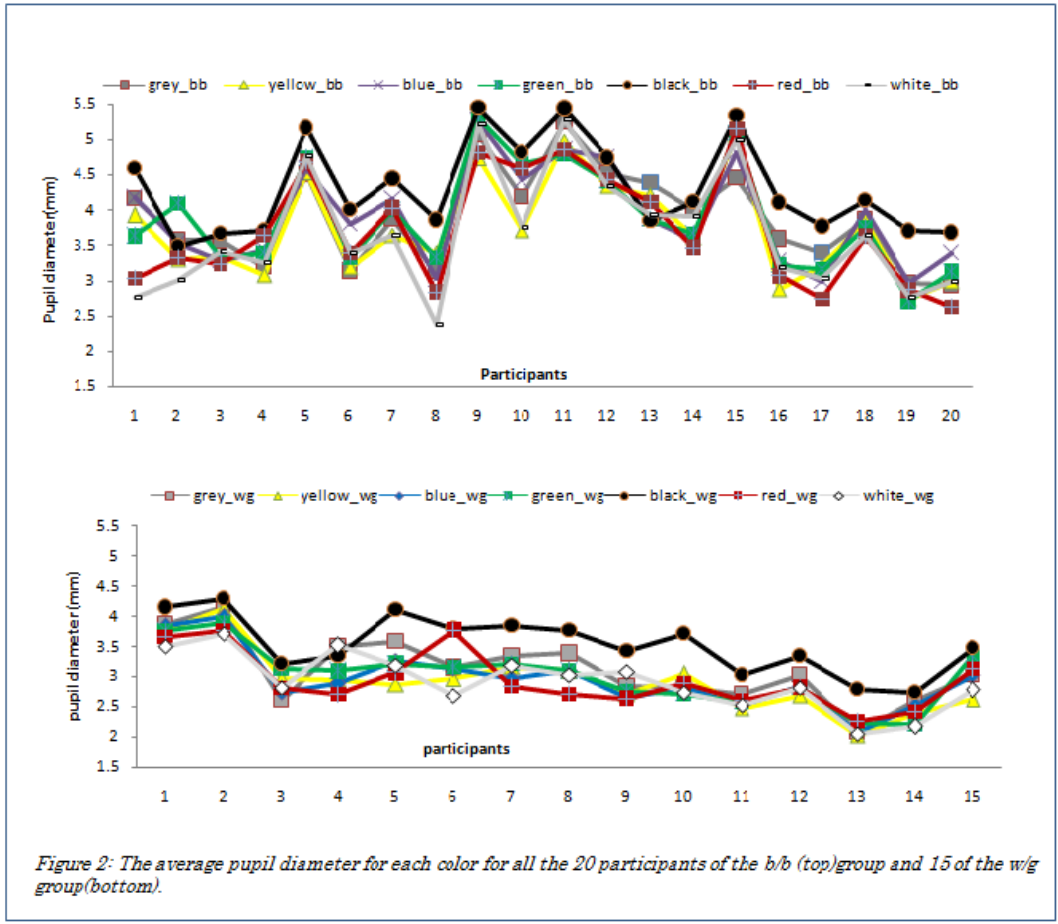

Figure 2: The average pupil diameter for each color for all the 20 participants of the b/b (top)group and 15 of the w/g group(bottom).

Table 1: The p-values for 7 colors from a 2-tailed unequal variance T-Test on the pupil size for the 2 groups (wg, bb).

| Color | p-value(b/b, w/g) |
|---|---|
| GREY | 0.000710167 |
| YELLOW | 0.000407803 |
| BLUE | 0.00004.12045 |
| GREEN | 0.000192334 |
| BLACK | 0.000330135 |
| RED | 0.000957234 |
| WHITE | 0.002005474 |

**Infinity focus**

The difference in pupil size between the two groups evidenced lead us to check if there was a population bimodality and the #dress has exposed it and hence an 'infinity' focus experiment as done in regular eye tests was set up. The repeat participants from the first experiment were 6 for each group, while 12 were new with a total of 18 in each group. The average pupil size from 5 seconds of data while focused on a 'X' sign 4 meters from the participant is co-plotted in Figure 3 for both groups in same lighting conditions as the first experiment. . The 2-sample T-test at unequal variance between the two groups for the condition wg< bb, estimated a T-value = 2.343 and p-value of 0.0127.

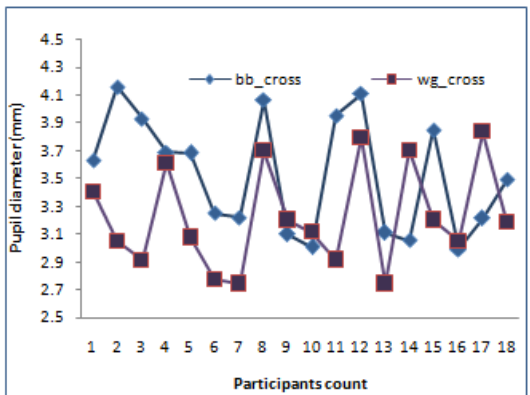

Figure 3: The pupil diameter variation for 18 subjects each of the two groups (w/g and b/b) collected as they were instructed to look at a projected 'X' sign on a screen 4 meters from the participant. The 'bb_cross' means b/b group and 'wg_cross' is w/g group.

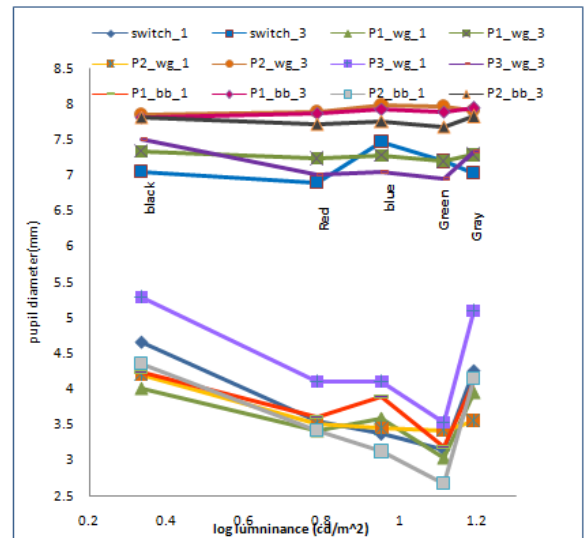

Figure 4: The pre-post dilation pupil response from 6 participants. The list on the left indicated the participant count (P1,P2, Switch) followed by the #dress group and the tag '1' is pre-dilation trail and '3' the second reading after dilation, which is 35 minutes after the pupil size increases.

### Artificial Pupil dilation

Would increase in pupil dilation artificially of the w/g especially bring about a change in color choice of the #dress? The hypothesis was prompted by the difference we noticed in pupil size between the b/b and w/g group. Six participants, 3 from each group, were administered dilation drops under the supervision of an ophthalmologist after all the required eye tests. One subject switched just before the pre-dilation test. A smaller slide set with RGB and black interspersed with 'gray' slides were shown for 3s and the #dress was shown once before and after the color slide show. The pupil size data collected from pre-dilation drops administration,  20 minutes into dilation state and 35 minutes later  show the expected inter-subject variation in the dilation size (pre-dilation:~3.3mm  to dilated state: 7.9mm) and except for one participant from the w/g group, larger pupil size was seen in b/b group in dilated state. The pupil size change as a function of the luminance (Figure 4) for the pre-post dilation for the six subjects show that in the dilated state the color response is almost flat for 5 participants expect for the 'switch' participant. But the color of the dress reported by the participants did not change from their initial choices, that is, the w/g group did not switch to b/b when dilated.

Our initial premise of a possible group-wise differential response to colors from the lower or higher wavelengths (Blue or red) was not observed but we found new and interesting results summarized as : a) The average pupil size across all the 20 participants who report black/blue is significant  higher  compared to the other two groups, which alludes to a possible bimodality in participant set, which has been highlighted by the current #dress experiment setup . The statistically analysis to check for significance also confirms the results.  b) No significant difference is noticed for the participants who either see the dress as white/gold or are able to switch. c) As expected the pupil size varies as a function of the color with relative difference between the base gray color and the consecutive color ranging from 0.2 to 0.6 mm and this was found to be approximately uniform across the groups. d)  the statistical significance test from the infinity_focus experiment confirms the findings from the color_slide experiment of a probable bimodal population and that normal pupil size is plausibly associated with the perceived color. The artificial dilation of the pupil did not bring the switch in color choice which suggests that light flux is not the main physiological parameter driving color identification differences. A possibility that needs to be explored is the S,L and M cone density differences in the two groups.  Another factor that requires further experiments is to check is by artificially constricting the pupil size the retinal illuminance in the b/b group is reduced and a switch to w/g is realized.  It was reported [5] that by constricting the pupil a few participants could see b/b and in our study  too a few of the w/g participants  reported that they could see b/b as the brightness of the screen was reduced by 60% and the b/b group could see the dress as 'w/g' at maximum brightness. The b/b group having consistently higher normal pupil dilation also indicates that macular pigment differences in these two groups need to be studied and also if subtle contrast sensitivity difference exists between these two groups

## 4. Summary
We have presented original results from three experiments that indicate a possible bimodality in normal pupil size within people, which could be a one of the factors that leads to the difference in the colors seen in the #dress. The statistically significant estimate of difference in normal pupil size for the two groups throws up more questions that require systematic experiments by increasing the sample size and by measuring cone density, astigmatism and macular pigmentation differences in the two groups.

## Acknowledgements
The authors sincerely thank the effort put in by Elvis Singhal, summer intern at the institute in collecting the preliminary data. Also thank all the participants and the ophthalmologist, Dr. Preetha, for the enthusiasm and interest in the experiment.

Links:
#dress: (http://swiked.tumblr.com/post/112073818575/guys-please-help-me-is-this-dresswhite-and)
Tobii: http://www.tobii.com/.